\documentclass[twocolumn,showpacs,prl,floatfix]{revtex4}

\newcommand{\mm}{\mathrm}

\ifx\pdfoutput\undefined
\usepackage{graphicx}
\else
\usepackage[pdftex]{graphicx}
\usepackage{epstopdf}
\fi

\usepackage[center]{subfigure}

\begin{document}

\title{Dislocations Jam At Any Density}

\author{Georgios Tsekenis, Nigel Goldenfeld and Karin A. Dahmen}
\affiliation{Department of Physics, University of Illinois at
Urbana-Champaign, Loomis Laboratory of Physics, 1110 West Green
Street, Urbana, Illinois, 61801-3080.}

\begin{abstract}
Crystalline materials deform in an intermittent way via
dislocation-slip avalanches. Below a critical stress, the dislocations
are jammed within their glide plane due to long-range elastic
interactions and the material exhibits plastic response, while above
this critical stress the dislocations are mobile (the unjammed phase)
and the material flows. 
We use dislocation dynamics and scaling
arguments in two dimensions to show that the critical stress grows with
the square root of the dislocation density. Consequently, dislocations
jam at any density, in contrast to granular materials, which only jam
below a critical density.
\end{abstract}


 \maketitle

When a crystalline material is sufficiently deformed, it undergoes
irreversible, or plastic deformation. Traditionally, plastic
deformation of crystalline solids has been considered to be a smooth
process in time, and homogeneous in space, since fluctuations are
expected to average out at sufficiently large spatial scales.  On
small spatial scales, however, intermittent motion of dislocations is
observed, resulting in a pattern of deformation that is spatially
inhomogeneous and intermittent in time.  This behavior is associated
with a coherent motion of the dislocations that releases stress by slip
avalanches: sequences of events with long-range correlations in space
and in time. The slip avalanches span several orders of magnitude in
size and the energy released is  distributed according to a power law
\cite{MiguelNat01, Richeton06, Richeton05, Weiss00, Weiss97,
DimidukSci06}.

Theoretical models, including discrete dislocation dynamics models
\cite{MiguelNat01, Zaiser1, Laurson06, CsikorSci07, Weygand2010, ZaiserAdvPhys06}, continuum models
\cite{Zaiser1,Zaiser2,Zaiser3, ZaiserAdvPhys06}, phase field models \cite{Koslowski, ZaiserAdvPhys06} and
phase field crystal models \cite{chanPRL10} are able to reproduce many
of the experimental findings and reveal scale invariant, power-law
distributed phenomena that are indicative of a non-equilibrium critical
point \cite{Sethna01}. The dislocation system is jammed below a
critical value of the external stress. Applying a constant external
stress above the critical (yield) stress allows the system to flow, and
the dislocations are unjammed. It is important to stress that in the
glide plane of the dislocations, there is effectively no external
potential, so that the jamming is an emergent phenomenon.  However,
recent work has shown that the behavior of the transition appears to be
in the universality class of the interface pinning-depinning transition
\cite{ZaiserAdvPhys06, chanPRL10}, as if there was an effective external potential
induced by the collective interactions between the dislocations.

\begin{figure}[h]
\begin{center}
\begin{tabular}{c}
\includegraphics[width=0.88\columnwidth]{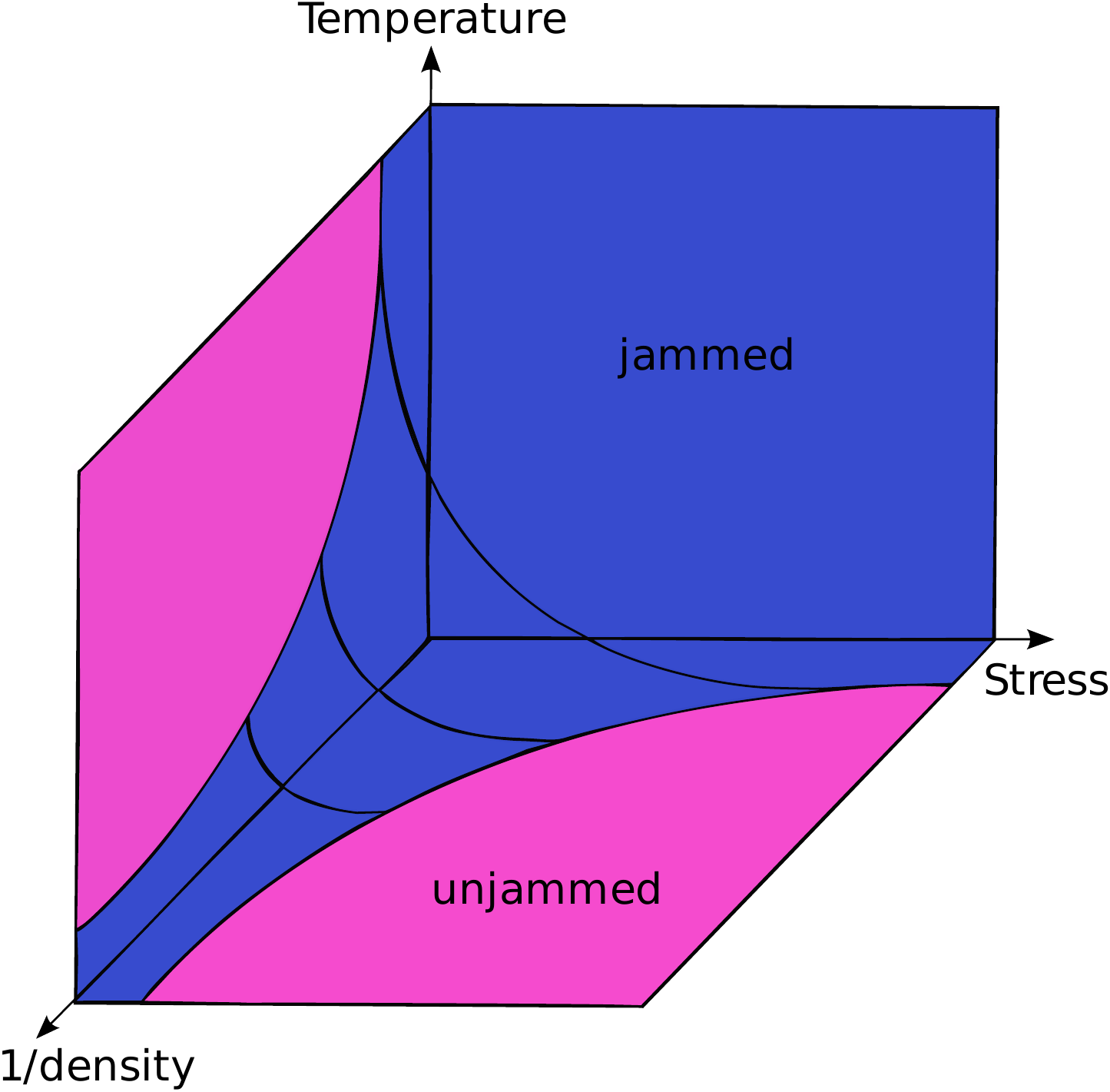}\\
\end{tabular}
\end{center}
\caption{(color online) Proposed phase diagram for dislocation systems. Notice the
absence of a jamming point.} \label{jammingPT}
\end{figure}

In this Letter we study connections between the plastic yield point
of systems with long range interactions, such as dislocation systems
and the jamming transition of systems with short range interactions, such
as sheared granular materials and molecular liquids
\cite{JammingNat98,JammingPRL02,JammingPRE03,LiuNagelRev2010}. When a
system jams it undergoes a transition from a flowing state (analogous
to a depinned phase) to a rigid state (analogous to a pinned phase). In
contrast to the ordered solid phase obtainable via crystallization, the
solid phase reached via jamming remains disordered. Liu and Nagel
\cite{JammingNat98}, O'Hern \textit{et al.}
\cite{JammingPRL02,JammingPRE03} and others \cite{LiuNagelRev2010}
studied jamming of granular materials with short range interactions in
simulations and experiments. They found that below a critical density
these materials do not jam at any stress. This critical density is
called the jamming point $J$ of granular materials. In contrast, we
show here that dislocations jam at any nonzero density, i.e.
dislocations have no jamming point. The physical
reason is that dislocations have long range interactions that can lead
to pinning for arbitrarily large distances between the dislocations.
Fig. 1 sketches the putative jamming phase diagram (in the absence of
screening) for dislocation-mediated plasticity. It is closely
related to the jamming phase diagram of \cite{LiuNagelRev2010} for
granular materials, except for the absence of a jamming point $J$ for
dislocations. 

In the following we employ analytical calculations and discrete
dislocation dynamics simulations to study how the critical yield stress
depends on the dislocation density $\rho$. Our analytical calculations
verify and generalize the numerical findings.

{\it The Model:-\/} We place $N$ straight edge dislocations parallel to
the $z$-axis in a square box of side $L$. They are allowed to glide only
along the shear direction ($x$-axis), while they can interact in the $x$
and $y$ directions. This simulates a single-slip system. Materials, like
ice, with strong plastic anisotropy, deform by glide on a single plane
\cite{MiguelNat01}. In these systems dislocation climb is negligible
due to high plastic anisotropy. The $z$ direction has been shown to be
irrelevant to scaling \cite{MiguelNat01,Laurson06, Zaiser1, CsikorSci07, Weygand2010}
effectivelly rendering the problem two-dimensional. An edge dislocation
with Burgers vector $\vec{b}=(b,0)$ produces in the host medium an
elastic shear stress at a distance $\vec{r}=(x,y)$,
\begin{eqnarray}
\tau_{\mm{int}}(\vec{r})=\frac{b\mu}{2\pi(1-\nu)} \frac{x(x^2-y^2)}{(x^2+y^2)^2}\label{tauint}
\end{eqnarray}
where $\mu$ is the shear modulus and $\nu$ is the Poisson ratio of the
host medium \cite{HirthLothe}. This is anisotropic in the (x,y)-plane
and decays as $\tau_{\mm{int}} \sim 1/r$ at large
$r=\sqrt{x^2+y^2}$. If an external shear stress
$\tau_{\mm{ext}}$ is applied the overdamped equation of motion of a
dislocation along the shear direction is described by,
\begin{eqnarray}
\eta \frac{dx_{i}}{dt} = b_{i} ( \sum^{N}_{j \neq i} \tau_{\mm{int}}(\vec{r}_{j}-\vec{r}_{i})-\tau_{\mm{ext}})
\end{eqnarray}
for $ i,j=1,...,N$ where $x_{i}$ is the $x$ coordinate of the $i$th
dislocation at point $\vec{r}_{i}$ with Burgers vector $b_i$,
$\vec{r}_{j}$ with $j \neq i$ are the coordinates of the other $N-1$
dislocations, $t$ is time and $\eta$ is the effective viscosity in the
host medium \cite{MiguelNat01,Laurson06,Zaiser1}. Here we have set the
temperature to $T=0$. For the computer simulations we have set the
distance scale $b=1$ and the time scale $t_{0}=\eta
/(\mu/(2\pi(1-\nu))=1$. To simulate bulk materials we employ periodic
boundary conditions in both $x$ and $y$ directions.

To treat the long-range character of the dislocation interaction, we
found the Lekner method \cite{Lekner} of image cells particularly
straight forward.
The equations of motion are solved by the adaptive-step fifth-order Runge-Kutta method \cite{NumRecC}.
The dislocation number is constant, since so far we considered neither
dislocation creation nor annihilation.
Equal numbers of dislocations with positive, $\vec{b}=+\hat{x}$, and
negative, $\vec{b}=-\hat{x}$, Burgers vectors, render the system
neutral. The dislocation collective speed (also called activity) $V(t)$, is defined as,
$V(t)=\sum^{N}_{i=1}|v_{i}(t)|$ where $v_{i}=dx_{i}/dt$. The acoustic emission signal is proportional
to the dislocation collective speed.
Another popular choice is $V(t)=\sum^{N}_{i=1}b_{i}v_{i}(t)$, which is proportional to the strain
rate \cite{ZaiserAdvPhys06}.

\begin{figure}[h]
\begin{center}
\begin{tabular}{c}
\includegraphics[width=0.88\columnwidth]{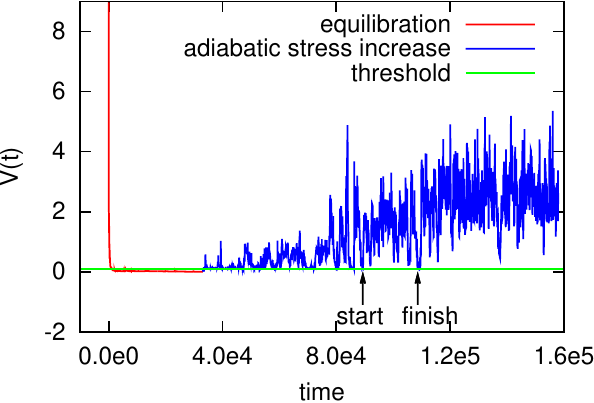}\\
\includegraphics[width=0.88\columnwidth]{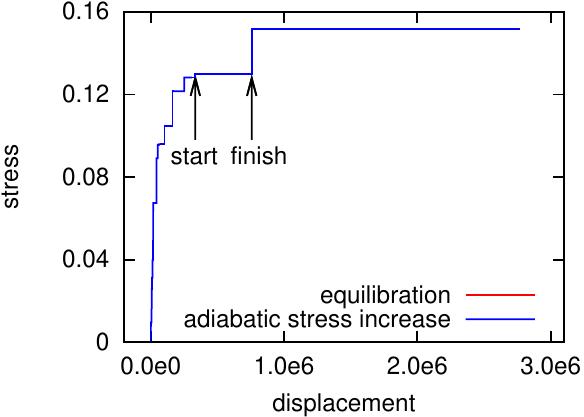}\\
\end{tabular}
\end{center}
\caption{(color online) (top) Time dependence of the collective speed
of all $N=64$ dislocations in a square box of side $L=100$.
(bottom) The stress plotted against the total dislocation displacement for the same run. 
Displacement at time $t$ is the total distance all the dislocations traveled
from the beginning of the simulation ($t=0$) till time $t$: $\int^{t}_{0} dt' \sum^{N}_{i=1}b_i dx_i (t')$.
The arrows indicate
the start and finish of the last large avalanche.
(Note that in the stress vs displacement bottom
figure the equilibration occurs at zero external stress.)}
\label{VvsTime}
\end{figure}

{\it Adiabatic Increase of External Stress:-\/} First we consider the
quasi-static or adiabatic case. After randomly placing the dislocations
in the square cell, we let the system relax to the nearest (metastable)
equilibrium state. During that procedure we apply zero external stress.
As the system approaches the nearest energy minimum the dislocation
motion slows down. A simple eigenmode analysis shows that the time
needed for the system to reach zero activity diverges. 
We assume that the system is
sufficiently close to the energy minimum when the dislocation activity
has fallen below a threshold, $V_{\mm{th}}=0.1$, which is roughly $100$
times less than the initial activity of the $N=64$ dislocation system.
Once the system's activity has fallen below the specified threshold we
start increasing the external stress adiabatically slowly. As soon as
the adiabatically slowly increased stress pushes the system's activity
above the threshold, $V(t)>V_{\mm{th}}$ and the system produces an
avalanche, we keep the external stress constant until the avalanche
stops (Fig. \ref{VvsTime}). 
The scaling behavior is insensitive to the threshold for a value
up to ten times larger and smaller.

\begin{figure}[h]
\begin{center}
\begin{tabular}{c}
\includegraphics[width=0.88\columnwidth]{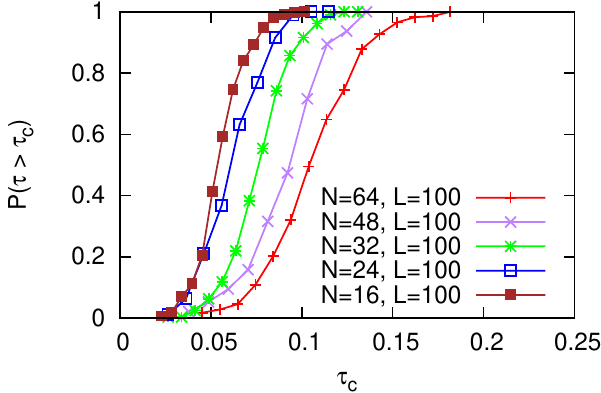}\\
\includegraphics[width=0.88\columnwidth]{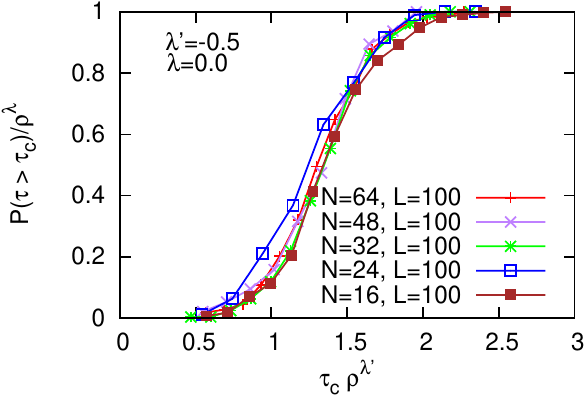}\\
\end{tabular}
\end{center}
\caption{(color online) (top) The cumulative distribution of critical stresses
$\tau_c$ for 5 different numerical densities. Each curve is extracted
from 288 runs with $N=64,32,16$ dislocations and 96 runs with $N=48,24$
dislocations in a square box of side $L=100$. The smaller the density,
the narrower the distribution and smaller the mean $\tau_c$ (see next
Fig. \ref{tau1overd}). (bottom) We obtain a good collapse using the
curves with the three larger densities based on the expression
$p(\tau_{c},\rho) \sim \rho^{\lambda}f[\tau_{c}\rho^{\lambda'}]$. The
collapse quantifies the fact that the distributions get steeper and
have a smaller mean for lower density $\rho$. $\lambda=0$ since
the cumulative probability is restricted in $[0,1]$. $\lambda'=-0.5 \pm 0.02$. 
The rescaling of the horizontal axis indicates that
$\tau_{c} \sim \rho^{0.5}$.}\label{distTausNallL100cum}
\end{figure}

The system starts with small avalanches and as the stress $\tau$
approaches the flow 
 stress $\tau_c$, it responds with larger and
larger avalanches until at $\tau_c$ it finally flows steadily 
with an infinite
avalanche. For $\tau > \tau_c$, the dislocations keep moving
indefinitely, exiting from one side of the simulation cell and
reemerging at the other due to the periodic boundary conditions,
without ever getting jammed (pinned) again. In a deformation experiment, this is
the point when the sample yields. In summary, for $\tau < \tau_c$ the
system is jammed (pinned).
For $\tau > \tau_c$ the system is
constantly flowing (yielding) (Fig. \ref{VvsTime}).

{\it Jamming:-\/} The critical stress $\tau_{c}$ is not a universal
quantity and every system with the same number of dislocations and box
size has a different $\tau_{c}$. We performed an adiabatically slow
increase of the stress for different dislocation densities,
$\rho=N/L^{2}$. The cumulative distributions of the critical stresses
is shown in top Fig. \ref{distTausNallL100cum}. One can observe that the
distributions become narrower for smaller densities, as does the mean
critical stress of the ensemble. The scaling collapse shown in
bottom Fig. \ref{distTausNallL100cum} gives the relationship $\tau_{c} \sim \sqrt{\rho}$.

\begin{figure}[h]
\begin{center}
\begin{tabular}{c}
\includegraphics[width=0.88\columnwidth]{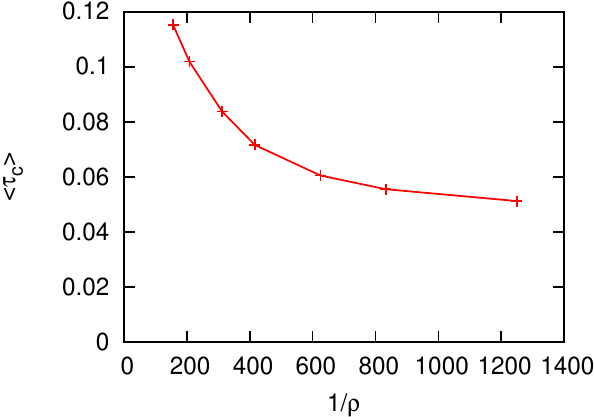}\\
\end{tabular}
\end{center}
\caption{(color online) The mean critical stress $\langle \tau_c \rangle$ plotted
against the inverse numerical density $\rho$. The system is jammed for
$\tau < \tau_c$ and unjammed above it.
Each point is extracted from 288 runs with $N=64,32,16,8$ dislocations and 96 runs with $N=48,24,12$ dislocations in a square box of side $L=100$.
The critical stress has a
similar qualitative dependence on the density as in the proposed
jamming phase diagram by Liu and Nagel \cite{JammingNat98}. However for
dislocations $\tau_c(\rho)>0$ for all non-zero densities $\rho$.}
\label{tau1overd}
\end{figure}

The dislocation system exhibits jamming for $\tau < \tau_c$ analogous
to the work of Liu and Nagel  \cite{JammingNat98} and O'Hern \textit{et
al.} \cite{JammingPRL02,JammingPRE03}. Their systems are different from
ours in that they had exclusively short range interactions (contact
interactions of soft spheres) and we have long-range (besides the core
interactions that are enforcing the "no climb" constraint). They
observed similar distributions of depinning stresses and a similar
concave up dependence of the flow stress on the density (Fig. \ref{tau1overd}). However in
contrast to their results, we neither expect nor find a
jamming point equivalent to their jamming point \textit{J} where
$\tau_c=0$. This means there can be no density, however small, that
will unjam our system at zero applied external stress. The reason is
that dislocations have long-range interactions \cite{LiuNagelRev2010}.
No matter how far apart they are, they always feel each other.

{\it Theoretical Calculation of Critical Stress:-\/} Consider
$N^{+}_{R,\Delta R}$ positive and $N^{-}_{R,\Delta R}$ negative edge
dislocations parallel to the $z$-axis randomly distributed on a ring of
radius $R$ and thickness $\Delta R$ on the $(x,y)$-plane. The stress
exerted at the origin is given by,
\begin{eqnarray}
\tau_{R,\Delta R} = \int^{R+\Delta R}_{R}{d^{2}r \frac{\rho^{+}(r,\theta)-\rho^{-}(r,\theta)}{r}K(\theta)}
\end{eqnarray}
adapted from \cite{GromaBakoPRB98} using Eq.(\ref{tauint}) where
$K(\theta) \sim \cos(\theta)\cos(2\theta)$ and
$\rho^{\pm}(r,\theta)=\sum^{N^{\pm}_{R,\Delta
R}}_{i=1}{\frac{\delta(r-r_i)}{r^{d-1}}\delta(\theta-\theta_i)}$. We
express all distances in terms of $l$, the mean dislocation distance,
i.e. $\rho=N/L^d=1/l^d$ in $d$ dimensions, i.e. $X=R/l$ and $x=r/l$.
For any power law $r^{-\alpha}$ interaction, we get $\tau_{X,\Delta X}
= \frac{l^d}{l^{\alpha}}\int^{X+\Delta X}_{X}{d^{d}x
\frac{\rho^{+}(x,\theta)-\rho^{-}(x,\theta)}{x^{\alpha}}K(\theta)}$
with $\rho^{\pm}(x,\theta)=\frac{1}{l^d}\sum^{N^{\pm}_{X,\Delta
X}}_{i=1}{\frac{\delta(x-x_i)}{x^{d-1}}\delta(\theta-\theta_i)}$. For
small ring  thickness we can approximate the integral with the value of
the integrand at $X$ times $\Delta X$. The average over the number of
the dislocations to first order gives
\begin{eqnarray}
\langle \tau_{X,\Delta X} \rangle\sim \frac{1}{l^{\alpha}}\frac{\Delta
X}{X^{\alpha}}(\langle N^{+}_{X,\Delta X} \rangle - \langle
N^{-}_{X,\Delta X}\rangle)=0
\end{eqnarray}
with $\langle N^{+} \rangle=\langle N^{-}
\rangle=\langle N \rangle$. The effect of the number fluctuations on
the stress per ring thickness is:
\begin{eqnarray}
\langle(\frac{\tau_{X,\Delta X}}{\Delta X})^2\rangle \sim \frac{\langle(N^{+}_{X,\Delta X}-N^{-}_{X,\Delta
X})^2\rangle}{l^{2\alpha} X^{2\alpha}} \sim \frac{1}{l^{2\alpha}}\frac{\langle
N_{X,\Delta X}\rangle}{X^{2\alpha}}\label{taufluc}
\end{eqnarray}
since $N^{\pm}$ are independent random variables, Poisson distributed
with the same mean and variance. Assuming that there are $N$
dislocations of each kind in the entire area $L^d$ where
$X_{L}=L/l>>1$, their mean number in the ring can be expressed as
$\langle N_{X,\Delta X}\rangle \sim N \frac{ X^{d-1} \Delta
X}{X_{L}^d}$. Substituting into Eq.(\ref{taufluc}) we
find, $\langle(\frac{\tau_{X,\Delta X}}{\Delta X})^2\rangle \sim \frac{1}{l^{2\alpha}} \frac{N}{X_{L}^d}\frac{\Delta X}{X^{2\alpha-d+1}}$
Integrating over the entire region,
\begin{eqnarray}
\frac{\tau^2}{X_{L}^d} \equiv \int{\langle(\frac{\tau_{X,\Delta X}}{\Delta X})^2\rangle} \sim \frac{1}{l^{2\alpha}} \frac{N}{X_{L}^d}\int^{X_{L}}_{X_{min}}{\frac{dX}{X^{2\alpha-d+1}}}
\end{eqnarray}
with $X_{min} \sim O(1)$ the closest possible distance between 2 dislocations gives: $\frac{\tau}{\sqrt{N}} \sim \frac{1}{l^{\alpha}}\frac{1}{\sqrt{2\alpha-d}}\sqrt{\frac{1}{X_{min}^{2\alpha-d}}-\frac{1}{X_{L}^{2\alpha-d}}}$ for $2\alpha>d$. In the thermodynamic limit, $X_{L} \to \infty$, this translates to the stress scaling as $\frac{\tau}{\sqrt{N}} \sim \frac{1}{l^{\alpha}} \sim \rho^{\frac{\alpha}{d}}$. For $2\alpha<d$, $\frac{\tau}{\sqrt{N}} \sim \frac{1}{l^{\alpha}}\frac{1}{\sqrt{d-2\alpha}}\sqrt{X_{L}^{d-2\alpha}-X_{min}^{d-2\alpha}}$ and  the thermodynamic limit doesn't exist. For parallel straight edge dislocations in 2 dimensions $2\alpha=d=2$ and
\begin{eqnarray}
\frac{\tau}{\sqrt{N}} \sim \frac{1}{l}\sqrt{\ln(L/l)} \sim \sqrt{\rho}\sqrt{\ln(L/l)}\label{taufluc2}
\end{eqnarray}
This agrees with our numerical result in
bottom Fig. \ref{distTausNallL100cum}. The extraction of logarithmic
corrections requires much larger systems than the ones that can be simulated.

{\it Discussion:-\/} We were able to show, using a discrete dislocation
dynamics model, that the mean critical stress of an ensemble of
dislocation systems with long-range interactions, $\tau_{\mm{int}} \sim
1/r$, scales with the square root of the dislocation density, $\langle
\tau_{c} \rangle \sim \sqrt \rho$, for straight parallel edge
dislocations. Eq. (\ref{taufluc2}) also agrees with the Taylor hardening relation \cite{HirthLothe} 
and is analogous to the effective velocity of a point vortex in 2 dimensional hydrodynamics \cite{ChavanisPRE02}. 
We were able to perform the analytical calculation for any power law
interaction, $\tau_{\mm{int}} \sim 1/r^{\alpha}$, and for arbitrary $d$
dimensions. The theoretical result agrees with our simulation up to
logarithmic corrections which are difficult to measure at system sizes
amenable to simulation. Our results, both numerical and theoretical,
show that for dislocations or particles with long-range interactions
there can be no jamming point at a finite density (only at $\rho=0$),
provided there is no screening.

{\it Acknowledgements:} We thank
M. -C. Miguel, M. Zaiser, J. Weiss, S. Zapperi, L. Laurson, M. Alava,
D. Ceperley, V. Paschalidis, K. Schulten, R. Brunner, J. Estabrook,
 J. T. Uhl, Y. Ben-Zion, S. Papanikolaou, J. Sethna, B. Brinkman, 
 L. Angheluta, H. Jaeger, A. Liu and S. Nagel
for helpful conversations. We acknowledge NSF grant DMR
03-25939 ITR (MCC) and DMR 1005209, the University of Illinois Turing cluster,
the KITP at UCSB (KD) and NSF grant TG-DMR090061 for TeraGrid TACC and NCSA resources.


\bibliographystyle{apsrev}

\bibliography{DDD_bib}

\end{document}